\documentclass[aps,nofootinbib,showpacs]{revtex4}
\usepackage{graphicx}
\usepackage{dcolumn}
\usepackage{bm}
\usepackage[dvips]{color}

\begin{document}

\hfill{LAPTH-1033/04, IFIC/04-11, PCC-0413}

\title{Probing neutrino masses with future galaxy redshift surveys}

\author{Julien Lesgourgues}
\affiliation{Laboratoire de Physique Th\'eorique LAPTH
(CNRS-Universit\'e de Savoie), B.P. 110,
 F-74941 Annecy-le-Vieux Cedex, France}
\author{Sergio Pastor}
\affiliation{Instituto de F\'{\i}sica Corpuscular (CSIC-Universitat de
Val\`encia), Ed.\ Institutos de Investigaci\'on, Apdo.\ 22085,
E-46071 Valencia, Spain}
\author{Laurence Perotto}
\affiliation{Physique Corpusculaire et Cosmologie (CNRS-IN2P3), 
11 place Marcelin Berthelot, 75231 Paris Cedex 05, France} 
\date{\today}

\begin{abstract}
    We perform a new study of future sensitivities of galaxy redshift
    surveys to the free-streaming effect caused by neutrino masses,
    adding the information on cosmological parameters from
    measurements of primary anisotropies of the cosmic microwave
    background (CMB). Our reference cosmological scenario has nine
    parameters and three different neutrino masses, with a hierarchy
    imposed by oscillation experiments.  Within the present decade,
    the combination of the Sloan Digital Sky Survey (SDSS) and CMB
    data from the PLANCK experiment will have a 2$\sigma$ detection
    threshold on the total neutrino mass close to $0.2$ eV.  This
    estimate is robust against the inclusion of extra free parameters
    in the reference cosmological model. On a longer term, the next
    generation of experiments may reach values of order $\sum m_{\nu}
    = 0.1$ eV at 2$\sigma$, or better if a galaxy redshift survey
    significantly larger than SDSS is completed.  We also discuss how
    the small changes on the free-streaming scales in the normal and
    inverted hierarchy schemes are translated into the expected errors
    from future cosmological data.
\end{abstract}
\pacs{14.60.Pq, 95.35.+d, 98.80.Es}

\maketitle

\section{Introduction}

Neutrino physics has provided the first clear indication of particle
physics beyond the Standard Model, since we have experimental
evidences for non-zero neutrino masses. Analyses of data from
atmospheric and solar neutrino experiments have shown the allowed
regions for the squared mass differences ($\Delta m_\nu^2$) at two
different scales. Such values will be known with better precision in
the next years, in particular for the larger atmospheric $\Delta
m_\nu^2$ using the results of future long-baseline oscillation
experiments.

However, from oscillation experiments no information can be obtained
on the absolute values of neutrino masses, since the lightest neutrino
mass remains unconstrained. Tritium decay experiments tell us that
each neutrino mass cannot be larger than 2.2 eV (95\% CL) at present
\cite{Bonn:tw}, to be improved to $\sim$0.35 eV with KATRIN
\cite{Osipowicz:2001sq}. More stringent bounds exist from experiments
searching for neutrinoless double beta decay, that will be improved
in the near future \cite{Elliott:2002xe}, but unfortunately they
depend on the details of the neutrino mixing matrix.

Cosmology offers several advantages: the cosmic neutrino background
provides an abundant density of relic neutrinos with an equal momentum
distribution for all flavors (up to 1\% corrections), which implies that
mixing angles have no effect. Although neutrinos cannot be the
dominant dark matter component, they can still constitute a small, hot
part of the matter density producing an erasure of perturbations
at small scales through their free-streaming effect (for a review, see
e.g.\ \cite{Dolgov:2002wy}). A comparison with data from the large
scale structure (LSS) of the Universe is thus sensitive to neutrino
masses, as emphasized in \cite{Hu:1997mj}.

At present, cosmological data allow us to bound the total neutrino mass
to values of $\sum m_\nu \alt 0.6-1.0$ eV
\cite{Spergel:2003cb,Hannestad:2003xv,Elgaroy:2003yh,Tegmark:2003ud,
Barger:2003vs,Hannestad:2003ye,Crotty:2004gm}, depending on the
data and priors used. These ranges already compromise the 4 neutrino
scenarios that could explain the additional large neutrino mass
difference required by the LSND results (that also imply a fourth,
sterile neutrino), but is not yet capable of reaching the necessary
$0.1$ eV range in order to test the hierarchical 3 neutrino schemes.
But such small masses could be detected in the next future when more
precise cosmological data are available, in a parallel effort to those
of beta and double beta decay experiments on Earth.

In this paper we analyze the future sensitivities of cosmological data
to neutrino masses, extending the pioneering work \cite{Hu:1997mj} and
in particular the detailed analysis in \cite{Eisenstein:1998hr} (see
also \cite{Lesgourgues:1999ej}), that was more recently updated in
\cite{Hannestad:2002cn}. In contrast to this last work we consider, in
addition to ideal Cosmic Microwave Background (CMB) observations
limited only by cosmic variance, the experimental specifications of
satellite missions such as PLANCK and the mission concept CMBpol
(Inflation Probe), as well as ground-based detectors such as ACT or
SPTpol, that will extend the PLANCK data to smaller angular scales.
We also increase the number of cosmological parameters of previous
analyses, including also the helium fraction, extra relativistic
degrees of freedom, spatial curvature, dark energy with constant
equation of state, or a primordial spectrum with running
tilt. Finally, our work is the first one in which it is assumed that
neutrinos have three different masses, in order to compute accurately
the free-streaming effect associated to the mass schemes allowed by
oscillation experiments.

Note that throughout this work, we will assume that the LSS power
spectrum is measured solely with galaxy redshift surveys. For
complementary constraints based on gravitational lensing, we refer the
reader to Refs.\ \cite{Abazajian:2002ck,Kaplinghat:2003bh}.

This paper is organized as follows. In Sec.\ II we review the expected
values of neutrino masses and their impact on Cosmology. We describe
future CMB experiments and galaxy surveys in Sec.\ III and the method
to forecast the errors on cosmological parameters in Sec.\ IV. Finally,
we present our results in Sec.\ V, with a summary and conclusions in
Sec.\ VI.

\section{Neutrino masses}
\label{sec:numasses}

Nowadays we have experimental evidences for neutrino oscillations from
solar and atmospheric neutrino detectors, recently also supported from
data on neutrinos from artificial sources (Kamland and K2K). Detailed
analyses of the experimental data lead to the following values of the
mass squared differences (best fit values $\pm$ 3$\sigma$ ranges)
\begin{eqnarray}
\Delta m_{\rm atm}^2 = \Delta m_{32}^2 = 
(2.6^{+1.1}_{-1.2}) \times 10^{-3} ~{\rm eV}^2 \nonumber\\
\Delta m_{\rm sun}^2 = \Delta m_{21}^2 = 
(6.9^{+2.6}_{-1.5}) \times 10^{-5} ~{\rm eV}^2
\label{dm2values}
\end{eqnarray}
taken from \cite{Maltoni:2003da}. These ranges are only slightly
different in other recent analyses, see e.g.\ 
\cite{Fogli:2003th,Gonzalez-Garcia:2003qf}, while a lower $\Delta
m_{\rm atm}^2$ seems required by new Super-Kamiokande data and
3-dimensional atmospheric fluxes. The errors in the above equation
will be significantly reduced with new data from Kamland in the case
of $\Delta m_{21}^2$, and with data from future long-baseline
oscillation experiments such as MINOS, ICARUS and OPERA, which will
give the atmospheric $\Delta m^2$ with $10\%$ accuracy (reduced to
$5\%$ with the superbeam proposal JPARC-SK) \cite{Huber:2004ug}.
Current data also provide the allowed ranges of the neutrino mixing
angles $\theta_{12}$ and $\theta_{23}$, and an upper bound on
$\theta_{13}$.

Indications for a third, heavier $\Delta m_\nu^2$ exist from the LSND
experiment \cite{Aguilar:2001ty}, implying a fourth (sterile)
neutrino. Such a mass is already being tested by present cosmological
data, although not ruled out yet \cite{Crotty:2004gm,Hannestad:2003ye,
  Hannestad:2003xv,Elgaroy:2003yh}, and the LSND results will be
checked by the ongoing experiment MiniBoone. Here we choose not to
include such a large $\Delta m^2_\nu$ and consider only the values in
Eq.\ \ref{dm2values}.

\begin{figure}
\includegraphics[width=.45\textwidth]{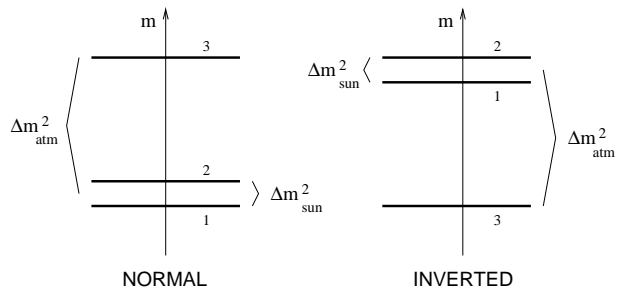}
\caption{\label{nusch}
The two neutrino schemes allowed if $\Delta m_{\rm atm}^2\gg
\Delta m_{\rm sun}^2$: normal hierarchy (NH) and inverted hierarchy (IH).}
\end{figure}

The three neutrino masses that lead to the values in Eq.\
\ref{dm2values} can be accommodated in two different neutrino schemes,
named normal ($m_3>m_2>m_1$) and inverted ($m_2>m_1>m_3$) hierarchy,
as shown in Fig.\ \ref{nusch}, that we will denote NH and IH. At
present we have no indication of which scheme is the correct
one. However, it has been suggested that some information could be
extracted from future data from Supernova neutrinos, very large
baseline oscillation experiments, or neutrinoless double beta decay
searches if the effective $m_\nu$ is below some threshold (for
reviews, see e.g.\ \cite{Gonzalez-Garcia:2002dz,Barger:2003qi}). In
general, determining the type of mass spectrum depends on the
precision with which the other mixing parameters would be measured.

Relic neutrinos were created in the Early Universe and decoupled from
the rest of the plasma when the temperature dropped below $\sim 1$
MeV, when they were ultra-relativistic. After decoupling all neutrino
flavors kept a Fermi-Dirac spectrum, only distorted at percent level
during the process of electron-positron annihilations into photons
\cite{Dolgov:1997mb,Mangano:2001iu}. It is well-known that massive
neutrinos could account for a significant fraction of the total energy
density of the Universe today, being their contribution directly
proportional to the number density. For vanishing neutrino chemical
potentials, the total neutrino contribution to the critical density is
given by
\begin{equation}
\Omega_\nu =  \frac{\sum m_\nu}{93.2~{\rm eV}} \, h^{-2}~,
\label{omeganuh2}
\end{equation}
where $h$ is the Hubble constant in units of $100$ km s$^{-1}$
Mpc$^{-1}$ and $\sum m_\nu$ runs over all neutrino mass states.  For
fixed neutrino masses, $\Omega_\nu$ would be enhanced if neutrinos
decoupled with a significant chemical potential (or equivalently, for
large relic neutrino asymmetries), but this possibility is now ruled
out \cite{Dolgov:2002ab}.

Therefore cosmology is at first order sensitive to the total neutrino
mass $\sum m_\nu=m_1+m_2+m_3$ (for the 3 neutrino schemes that we
consider), but blind to the neutrino mixing angles or possible CP
violating phases. This fact differentiates cosmology from terrestrial
experiments such as beta decay and neutrinoless double beta decay,
which are sensitive to $\sum_i | U_{ei}|^2 m_i^2$ and $|\sum_i
U_{ei}^2 m_i|$, respectively, where $U$ is the $3\times 3$ mixing
matrix that relates the weak and mass bases.

\begin{figure}
\includegraphics[width=.6\textwidth]{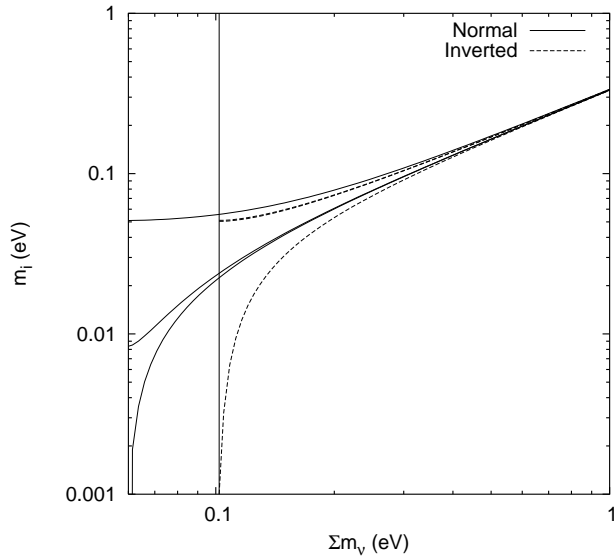}
\caption{\label{mi_vs_sum}
Neutrino masses as a function of the total mass in the two schemes
for the best-fit values of $\Delta m^2$ in Eq.\ \ref{dm2values}.
The vertical line marks the smallest value of $\sum m_\nu$ in
the inverted scenario.}
\end{figure}

It is interesting to see how the total mass is distributed among the
neutrino states for the two different schemes described above. They
are plotted in Fig.\ \ref{mi_vs_sum}. For a total mass above $\sim
0.2-0.3$ eV the two schemes are similar and correspond to a degenerate
scenario where each mass is $\sum m_\nu/3$.  However, for smaller
masses the number of neutrino states with relevant masses is 2 (1) in
the inverted (normal) hierarchy.

The effect of neutrino masses on cosmological observables has been
usually considered equivalent for fixed $\sum m_\nu$ (or $\Omega_\nu
h^2$). However, many papers noted in the past that this is not the
case and could potentially lead to differences, i.e.\ the neutrino
mass spectrum should be incorporated if the sensitivity to neutrino
masses is good enough (see, for instance the comments in
\cite{Lewis:2002nc,Hannestad:2002cn,Abazajian:2002ck}). As
an example, we note that in the mid-1990s it was shown that for CHDM
models with the same total neutrino mass (of order some eVs), those
with two degenerate massive neutrinos fitted better the data than
those with only one (see e.g.\ \cite{Primack:1994pe}).

\begin{figure}
\includegraphics[width=.48\textwidth]{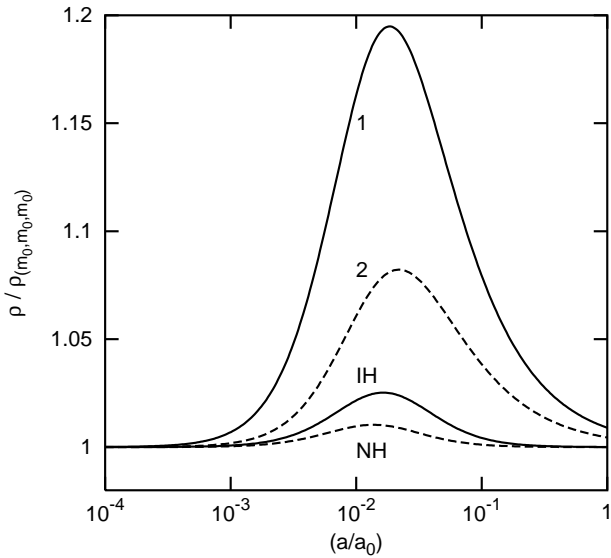}
\caption{\label{evolrho}
  Evolution of the total neutrino energy density as a function of the
  scale factor of the Universe for models where the same $\sum m_\nu$
  ($0.12$ eV) is distributed differently. Each line corresponds to the
  energy density of 4 different cases (only 1 or 2 massive states,
  Normal and Inverted Hierarchy) normalized to the case with 3 massive
  states with mass $m_0=\sum m_\nu/3$.}
\end{figure}

\begin{figure}
\includegraphics[width=.48\textwidth]{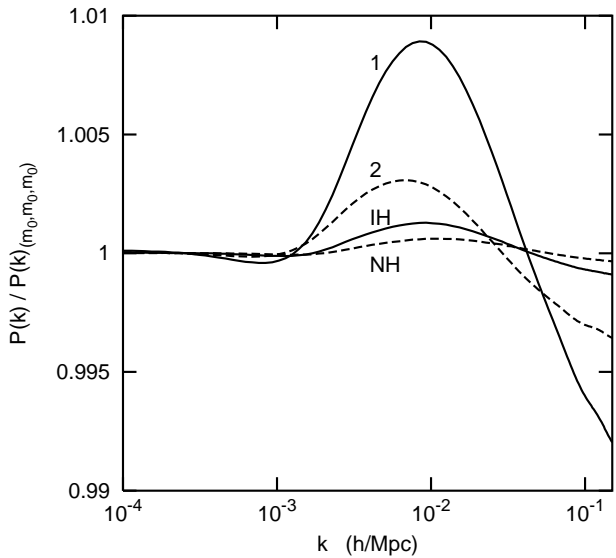}
\caption{\label{comparepk}
  Comparison of the matter power spectrum obtained for various models 
  where the same $\sum m_\nu$
  ($0.12$ eV) is distributed differently. The four lines correspond to
  the cases with 1 or 2 massive states,
  Normal and Inverted Hierarchy, divided each time by that with 3 massive
  states of equal mass $m_0=\sum m_\nu/3$. Differences in the various
  individual masses and free-streaming scales affect the position and
  amplitude of the break in the power spectrum.
}
\end{figure}

Fixed the total neutrino mass, a different distribution among the 3
states $(m_1,m_2,m_3)$ causes a slight modification of the transit
from a relativistic to a non-relativistic behavior. This can be seen
in Fig.\ \ref{evolrho}, where the evolution of the neutrino energy
density is plotted for several cases with the same total neutrino
mass, equally shared by 1,2 or 3 neutrino states, as well as the
realistic NH and IH schemes (taking the best-fit values of $\Delta
m^2$). Therefore, the evolution of background quantities is not
completely independent of the mass splitting. However, the main
difference appears at the level of perturbations. Indeed, in the case
of non-degenerate massive neutrinos, various free-streaming scales are
imprinted in the matter power spectrum $P(k)$. This is illustrated in
Fig.\ \ref{comparepk}, where we compare $P(k)$ in the same cases as in
Fig.\ \ref{evolrho}. These results were obtained with our modified
version of the public code CMBFAST \cite{Seljak:1996is} (see section
\ref{sec:results} for details).

We have recently summarized the effects of massive neutrinos on
cosmological observables in \cite{Crotty:2004gm}. Here we simply
remind that only neutrinos with masses close to the recombination
temperature ($T_{\rm dec} \sim 0.3$ eV) leave an imprint on the CMB
angular spectra, while neutrinos with smaller masses have almost the
same effect as massless neutrinos. On the other hand, the dominant
effect is the one induced by free-streaming on the matter power
spectrum.  Therefore, the usual strategy is to combine CMB and LSS
measurements, where the former roughly fix most of the cosmological
parameters, while the latter is more sensitive to neutrino masses.

\section{Future CMB and LSS data}
\label{sec:futureCMBandLSS}

In this section we briefly describe the experimental projects, planned
or in development, that will provide data on the CMB anisotropy spectrum
or on the distribution of LSS.

\subsection{CMB experiments}

The quality of the first-year data from the Wilkinson Microwave
Anisotropy Probe (WMAP) \cite{Spergel:2003cb}, complemented by the
results of other experiments at smaller angular scales such as ACBAR,
CBI or VSA \cite{Goldstein:2002gf,Readhead:2004gy,Rebolo:2004vp}, has
shown the importance of CMB data as a probe of cosmological
parameters.  The CMB experiments measure the temperature fluctuations
in the sky that can be expanded in spherical harmonics,
\begin{equation}
\frac{\Delta T}{T} (\theta,\phi) = \sum_{l,m} a_{lm}Y_{lm}(\theta,\phi)
\label{dT} \, .
\end{equation}
If the underlying perturbations are Gaussian, all information is encoded in 
the angular power spectrum $C_l \equiv \langle |a_{lm}|^2\rangle$.
In addition the CMB experiments can be sensitive to polarization
anisotropies, that are expressed in terms of the angular spectra of
the E and B modes of polarization, as well as the temperature
polarization cross-correlation (TE) spectrum.

After WMAP, the next satellite mission will be
PLANCK\footnote{\tt http://www.rssd.esa.int/index.php?project=PLANCK}, to
be launched in 2007, whose experimental parameters are listed in Table
\ref{tableexp}. After a couple of years, it will provide CMB data more
precise than that of WMAP, in particular concerning polarization. We
also consider the CMBpol or Inflation Probe mission concept, presented
in the framework of NASA's Beyond Einstein Program\footnote{\tt
http://universe.gsfc.nasa.gov/program/inflation.html}.
This experiment would have better sensitivity than the limit
imposed by cosmic variance (up to $l \sim 2300$ for E-polarization,
even beyond for temperature).

In parallel to the satellite missions, there will be ground-based
experiments that will measure the CMB at smaller angular scales with
significantly smaller sky coverage but good sensitivities, such as
SPTpol\footnote{\tt http://astro.uchicago.edu/spt/} (in construction),
ACT\footnote{\tt http://www.hep.upenn.edu/$\sim$angelica/act/act.html}
(funded in January 2004), or QUaD \cite{Bowden:2003ub} (in
construction). As an example, we consider SPTpol with the
characteristics listed in Table \ref{tableexp}.

The observed power spectrum can be decomposed into primary
anisotropies, gravitational lensing distortions, and foreground
contamination. The central frequencies of CMB detectors are usually
chosen in order to minimize the foreground contribution. In addition,
by combining various frequencies, future experiments will have the
power to separate efficiently the CMB blackbody from the various
foregrounds contributions, even on small angular scales where the
latter start to be significant. It is possible to build models for the
foregrounds and to predict their impact on parameter extraction
\cite{Tegmark:1999ke,Prunet:1999fj,Bowden:2003ub}; this approach is
rather model-dependent, since the level of many foreground signals has
not yet been measured experimentally. Here, we will not enter into
such details. When dealing with PLANCK, we will employ only three
frequency channels from the high frequency instrument (HFI), making
the (usual) simplifying assumption that other channels will be used
for measuring the various foregrounds, and for cleaning accurately the
primary signal. We will do similar assumptions for SPTpol and CMBpol.
We will also speculate on the results of an ``ideal CMB experiment''
limited only by cosmic variance. Then, we will limit ourselves to
$l_{\rm max}=2500$ both for temperature and polarization, which
assumes an efficient method for foreground subtraction -- in
particular of point-like sources and dust -- but remains realistic (as
indicated by Fig.\ 7 in \cite{Tegmark:1999ke}). For the two satellite
experiments, we assume a sky coverage of $f_{\rm sky}=0.65$, which
represents a conservative estimate of the data fraction that will be
included in the analysis in order to avoid galactic foregrounds. For
the ``ideal CMB experiment'', we adopt the more optimistic value
$f_{\rm sky}=1$, assuming that all galactic
foregrounds can be subtracted (see e.g.\ the component separation
method described in \cite{Patanchon:2003dg}).

The issue of gravitational lensing distortion is subtle and
potentially very interesting. Since lensing is induced by large scale
structure, mainly on linear scales, this effect can be accurately
predicted for a given matter power spectrum. Therefore, if the
gravitational distortion of the CMB maps could be measured directly,
there would be an opportunity to extract the matter power spectrum
(and the underlying cosmological parameters) independently from
redshift surveys. A way of doing this is described in 
\cite{Hu:2001fb,Hu:2001kj,Okamoto:2003zw}, 
and has been already applied to future neutrino mass extraction by
\cite{Kaplinghat:2003bh}.  Here, we do not incorporate this method,
and assume that the matter power spectrum is measured only with
redshift surveys, leaving a combined analysis for the
future. Therefore, throughout the analysis, we will employ the
unlensed CMB power spectra\footnote{Note that including lensing
corrections is technically easy with CMBFAST. However, this would
introduce some correlations among different modes and scales 
that would artificially
lower the predicted errors on each cosmological parameters
\cite{Hu:2001fb,Kaplinghat:2003bh}.}. For the T, E and TE modes, lensing
distortions are subdominant. In contrast, for the B mode, lensing is
expected to dominate over the primary anisotropies at least on small
angular scales.  The angle above which lensing is subdominant
crucially depends on the tensor-to-scalar ratio, an inflationary
parameter which order of magnitude is still unknown. So, we follow a
conservative approach and not take the B mode into account.  This
amounts in assuming that the gravitational wave background generated
by inflation is small, so that the B mode gives no information on
primary anisotropies.

\begin{table}
\begin{ruledtabular}
\begin{tabular}{cccccc}
Experiment & $f_{\rm sky}$ & $\nu$ & $\theta_b$ & $\Delta_T$ & $\Delta_P$\\
\hline
PLANCK & 0.65 & 100 & 9.5' & 6.8 & 10.9\\
&  & 143 & 7.1' & 6.0 & 11.4\\
&  & 217 & 5.0' & 13.1 & 26.7\\
\hline
SPTpol & 0.1 & 217 & 0.9' & 12 & 17\\
\hline
CMBpol & 0.65 & 217 & 3.0' & 1 & 1.4\\
\end{tabular}
\end{ruledtabular}
\caption{\label{tableexp}
Experimental parameters of CMB projects: here $\theta_b$ measures the
width of the beam, $\Delta_{T,P}$ are the sensitivities per pixel in
$\mu$K, $\nu$ is the center frequency of the channels in GHz and
$f_{\rm sky}$ the observed fraction of the sky. For the PLANCK 100 GHz
channel, the value of $\Delta_P$ takes into account the recent design
with eight polarized bolometers.}
\end{table}

\subsection{Galaxy surveys}

The existing data on the distribution of galaxies at large scales come
from several galaxy surveys, of which the completed 2dF
survey\footnote{\tt http://www.mso.anu.edu.au/2dFGRS/} and the ongoing
Sloan Digital Sky Survey\footnote{\tt http://www.sdss.org} (SDSS) are the
largest. SDSS will complete its measurements in 2005.  The matter
power spectrum $P(k)$ can be reconstructed from the data, which gives
an opportunity to test the free-streaming effect of massive neutrinos.
However, the linear power spectrum is found modulo a biasing factor
$b^2$, which reflects the discrepancy between the total matter
fluctuations in the Universe, and those actually seen by the
instruments.  Here we assume that the bias parameter $b$ is
independent of the scale $k$.

An important point concerning LSS data is the non-linear clustering of
the smallest scales. The usual approach is to discard any information
above an effective cut-off wavenumber $k_{\rm max}$, while considering
results at lower $k$'s as a direct estimate of the linear
power spectrum. The cut-off value must be chosen with care:
if $k_{\rm max}$ is too small, we can lose a lot of information,
especially concerning the neutrino free-streaming scale. If 
$k_{\rm max}$ is too large, we can underestimate the error on cosmological
parameters, first by neglecting any theoretical uncertainty
in the quasi-linear corrections that could be applied to the spectrum, 
and second by ignoring the non-gaussianity
induced by non-linear evolution \cite{Tegmark:1997rp}.

Apart from $k_{\rm max}$, the important parameter characterizing the
galaxy survey is its effective volume in $k$ space, defined in
\cite{Tegmark:1997rp}. If the number density of objects in the survey
$n({\mathbf r})$ is roughly constant over the survey volume, and if
the observed power spectrum $P(k)$ is bigger than $1/n$ over the
scales of interest (i.e., from the turn-over scale in $P(k)$ up to
$k_{\rm max}$), the effective volume is equal to the actual volume of
the survey. This is a reasonable approximation for all the examples
that we will consider here. For instance, the SDSS the Bright Red
Galaxy (BRG) survey has an effective volume of roughly $V_{\rm eff}
\simeq 1~ ({\rm Gpc}/h)^3$ \cite{Eisenstein:1998hr} (which comes from
a sky coverage $f_{\rm sky}=0.25$ and a radial length of $1$ Gpc
$h^{-1}$).

Beyond SDSS, plans for larger surveys are under discussion. For
instance, we can mention the Large Synoptic Survey 
Telescope\footnote{\tt http://www.lsst.org} (LSST),
which in the future could cover the entire sky and at the same time be
capable of measuring fainter objects \cite{Tyson:2003kb}. LSST is
designed mainly for weak lensing observations. In order to map the
total matter distribution up to half the age of the Universe
(i.e., up to a redshift
$z \sim 0.8$ or a radial length $l \sim 2.3~{\rm Gpc}/h)$ in
a solid angle 30,000 deg$^2$ ($f_{\rm sky} \sim 0.75$), it could
measure $2\times10^8$ redshifts up to $z=1.5$. Inspired roughly
by these numbers, at the end of
this analysis, we will speculate
on the possibility to measure the
power spectrum in a effective volume as large as
$V_{\rm eff} = (4 \pi/3) f_{\rm
sky} l^3 \sim 40~({\rm Gpc}/h)^3$.

The mechanism of structure formation affects larger wavelengths at
later times. So, in order to measure the linear power spectrum on
small scales, it would be very useful to build high redshift galaxy
surveys. This is one of the main goals of the Kilo-Aperture Optical
Spectrograph (KAOS) proposal\footnote{\tt http://www.noao.edu/kaos}.
KAOS could build two catalogs centered around redshifts $z= 1$ and
$z=3$, corresponding roughly to $k_{\rm max} \sim 0.2~h$ Mpc$^{-1}$
and $k_{\rm max} \sim 0.48~h$ Mpc$^{-1}$ respectively, instead of
$k_{\rm max} \sim 0.1~h$ Mpc$^{-1}$ today (conservative values).  In
both catalogs, the number density would be such that $1/n \sim
P(k_{\rm max})$, and the effective volume of the two samples close to
$V_{\rm eff} \sim 0.5\, ({\rm Gpc}/h)^3$ and $V_{\rm eff} \sim 0.6\,
({\rm Gpc}/h)^3$ respectively \footnote{The characteristics of KAOS
are taken from the ``Purple Book'' available on-line at {\tt
http://www.noao.edu/kaos}.}. This experiment is designed mainly
for measuring the scale of baryonic oscillations, in order to
constrain dark energy.  However, we will see that it would be also
appropriate for improving constraints on the neutrino masses.

\section{Forecast of future bounds: Fisher matrix analysis}

Since the characteristics of future CMB experiments and galaxy surveys
are already known with some precision, it is possible to assume a
``fiducial'' model, i.e., a cosmological model that would yield the
best fit to the future data, and employ the Fisher matrix method to
forecast the error with which each parameter will be extracted. This
method has been widely used for many cosmological parameters, some of
them related to neutrinos. For instance, we can mention that forecast
analyses based on the Fisher matrix have shown that with future data
there will be a potential sensitivity to an effective number of
neutrinos of the order $\Delta N_{\rm eff}\sim 0.2$
\cite{Lopez:1998aq,Bowen:2001in,Bashinsky:2003tk}, a value that is
complementary to and will eventually improve the accuracy of
primordial nucleosynthesis results (see e.g.\ 
\cite{Cuoco:2003cu,Cyburt:2003fe}).

Starting with a set of parameters $x_i$ describing the fiducial model,
one can compute the angular power spectra of CMB temperature and
polarization anisotropies $C_l^X$, where $X=T,E,TE$.  Simultaneously,
one can derive the linear power spectrum of matter fluctuations
$P(k)$, expanded in Fourier space. The error $\delta x_i$ on each
parameter can be calculated from the reduced (dimensionless) Fisher
matrix $F_{ij}$, which has two terms.  The first one accounts for the
CMB experiment and is computed according to ref.\ 
\cite{Jungman:1995av}
\begin{equation} 
F_{ij} ({\rm CMB})=\sum_{l=2}^{l_{\rm max}} 
\sum_{X,Y} \frac{\partial C_l^X}{\partial \ln x_i}
{\rm Cov}^{-1} (C_l^X, C_l^Y) \frac{\partial C_l^Y}{\partial \ln x_j} \, ,
\label{fisher.matrix}
\end{equation}
where ${\rm Cov} (C_l^X, C_l^Y)$ is the covariance matrix of the
estimators of the corresponding CMB spectrum. For instance, the
$TT$ element is given by
\begin{equation}
{\rm Cov} (C_l^{T}, C_l^{T})= 
\frac{2}{(2l+1)f_{\rm sky}}~\left[C_l^{T}+ (
\sum_{\rm ch.} \omega_T B_l^2 )^{-1} \right]^2~.
\label{CovTT}
\end{equation}
Here, the first term arises from cosmic variance, while the second is
a function of the experimental parameters summed over the channels:
$B_l^2=\exp(-l(l+1)\theta_b^2/8\ln 2)$ is the beam window function
(assumed to be Gaussian), $\theta_b$ is the FWHM of the beam and
$\omega_T = (\theta_b \Delta_T)^{-2}$ is the inverse square of the
detector noise level ($\Delta_T$ is the sensitivity per pixel, and the
solid angle per pixel can be approximated by $\theta_b^2$).  For the
experiments that we consider here, all these numbers can be found in
Table \ref{tableexp}.  The other terms of the covariance matrix can be
found, for instance, in \cite{Eisenstein:1998hr}.

The second term of the reduced Fisher matrix accounts for the galaxy survey
data and is calculated following Tegmark \cite{Tegmark:1997rp},
\begin{equation}
F_{ij} ({\rm LSS})= 2 \pi \int_{0}^{k_{\rm max}}
\frac{\partial \ln P_{\rm obs} (k)}{\partial \ln x_i}
\frac{\partial \ln P_{\rm obs} (k)}{\partial \ln x_j} w(k) ~d \ln k.
\label{fisher.matrix2}
\end{equation}
Here $w(k)=V_{\rm eff}/(2\pi/k)^3$ is the weight function of the
galaxy survey and we have approximated the lower limit of the integral
$k_{\rm min} \simeq 0$. We defined $P_{\rm obs}(k) \equiv b^2 P(k)$, and
$k_{\rm max}$ is the maximal wave number on which linear predictions
are reliable. This expression is only an approximation, since in
addition to non-linear clustering it ignores edge effects and redshift
space distortions.

Inverting the total Fisher matrix, one obtains an estimate of the
1-$\sigma$ error on each parameter, assuming that all other parameters
are unknown
\begin{equation}
\frac{\delta x_i}{x_i} = (F^{-1})_{ii}^{1/2}.
\end{equation}
It is also useful to compute the eigenvectors of the reduced Fisher
matrix (i.e., the axes of the likelihood ellipsoid in the space of
relative errors). The error on each eigenvector is given by the
inverse square root of the corresponding eigenvalue. The eigenvectors
with large errors indicate directions of parameter degeneracy; those
with the smallest errors are the best constrained combinations of
parameters.

\section{Results}
\label{sec:results}

We have computed the total Fisher matrix from Eqs.\ 
\ref{fisher.matrix} and \ref{fisher.matrix2}, using various
experimental specifications.  Throughout the analysis, our fiducial
model is the concordance ``flat $\Lambda$CDM'' scenario, with
parameters close to the current best-fit values and with additional
neutrino masses. The nine free parameters with respect to which
derivatives are computed are: $\Omega_m h^2$ (matter density,
including baryons, cold dark matter and neutrinos), $\Omega_b h^2$
(baryon density), $\Omega_{\Lambda}$ (cosmological constant),
$C_{200}^T$ (amplitude of temperature spectrum at multipole 200),
$n_s$ (scalar tilt), $\tau$ (optical depth to reionization), $y_{He}$
(fraction of baryonic mass in the form of Helium), $M \equiv \sum
m_\nu$ (total neutrino mass) and $b$ (unknown bias of the LSS data).
The fiducial value of $b$ is irrelevant by construction, and we will
try various values of $M$, distributed following the NH or IH scheme.
Other fiducial values read:
$$
(\Omega_m h^2,\Omega_b h^2, \Omega_{\Lambda}, C_{200}^T, n_s, \tau, y_{He}) =
(0.143, 0.023, 0.70, 0.85, 0.96, 0.11, 0.24).
$$
All derivatives are computed at zero spatial curvature (by varying
$h$ appropriately). Note that we use double-sided derivatives with
step 10\% for $M$, 50 \% for $y_{He}$, 5\% for all other parameters.
We checked carefully that these steps are sufficient in order to avoid
possible numerical errors caused by the limited precision of the
Boltzmann code -- in our case, version 4.5.1 of CMBFAST
\cite{Seljak:1996is}, with option ``high precision''. We also checked
that with twice larger steps, the results change only by a negligible
amount. These conditions were not a priori obvious for the smallest
neutrino masses studied here, but we increased the precision of the
neutrino sector in CMBFAST accordingly. Actually, in order to study
three neutrino species with different masses, we performed significant
modifications throughout CMBFAST. For each mass eigenstate, we
integrate some independent background and perturbation equations,
decomposed in 15 momentum values, up to multipole $l = 7$. Finally,
we include the small distortions in the neutrino phase-space
distributions caused by non-instantaneous decoupling from the
electromagnetic plasma (with QED corrections at finite temperature)
\cite{Mangano:2001iu}, but these last effects are almost
negligible in practice.

\subsection{PLANCK+SDSS}

\begin{figure}
\includegraphics[width=.48\textwidth]{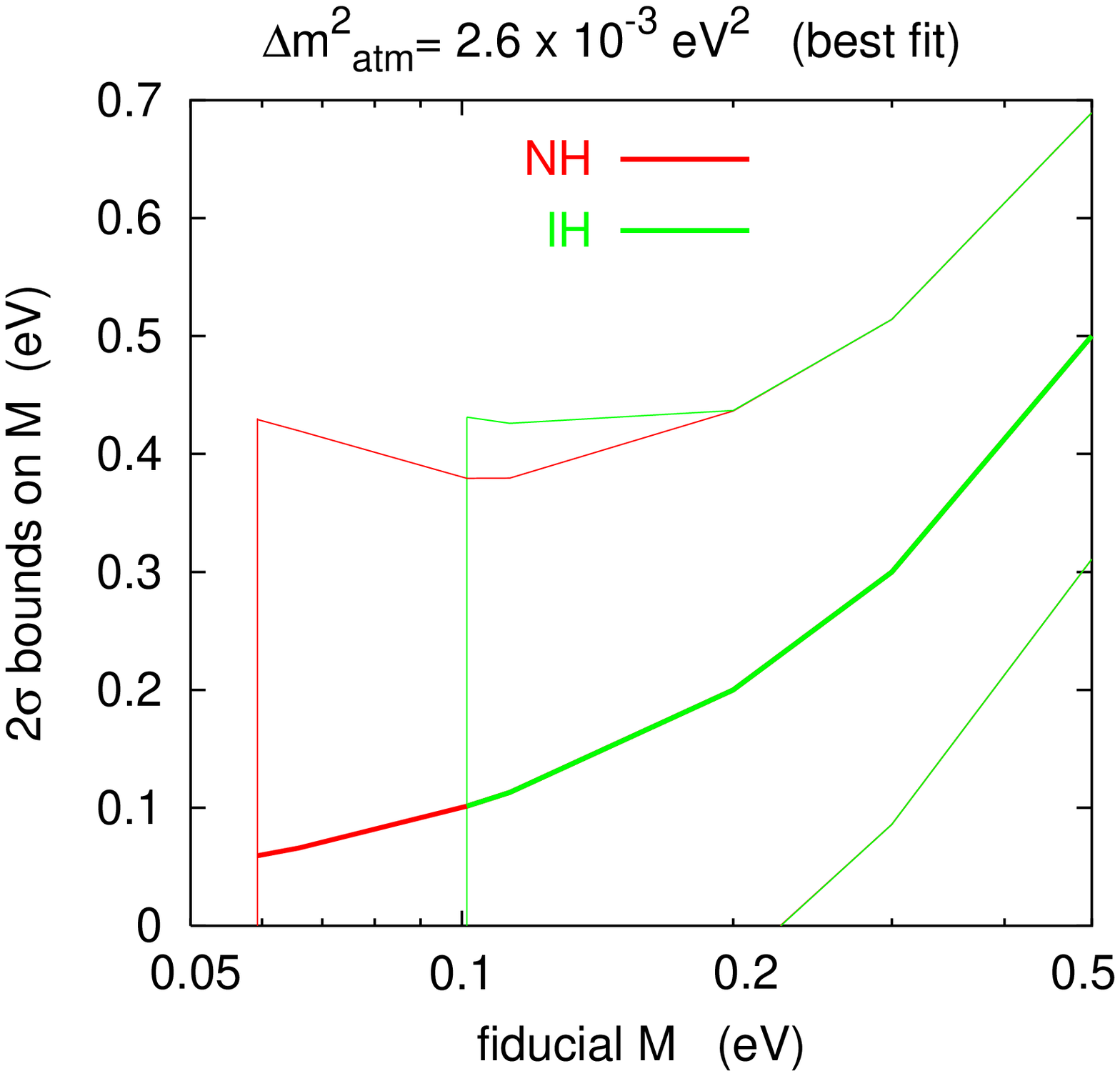}
\includegraphics[width=.48\textwidth]{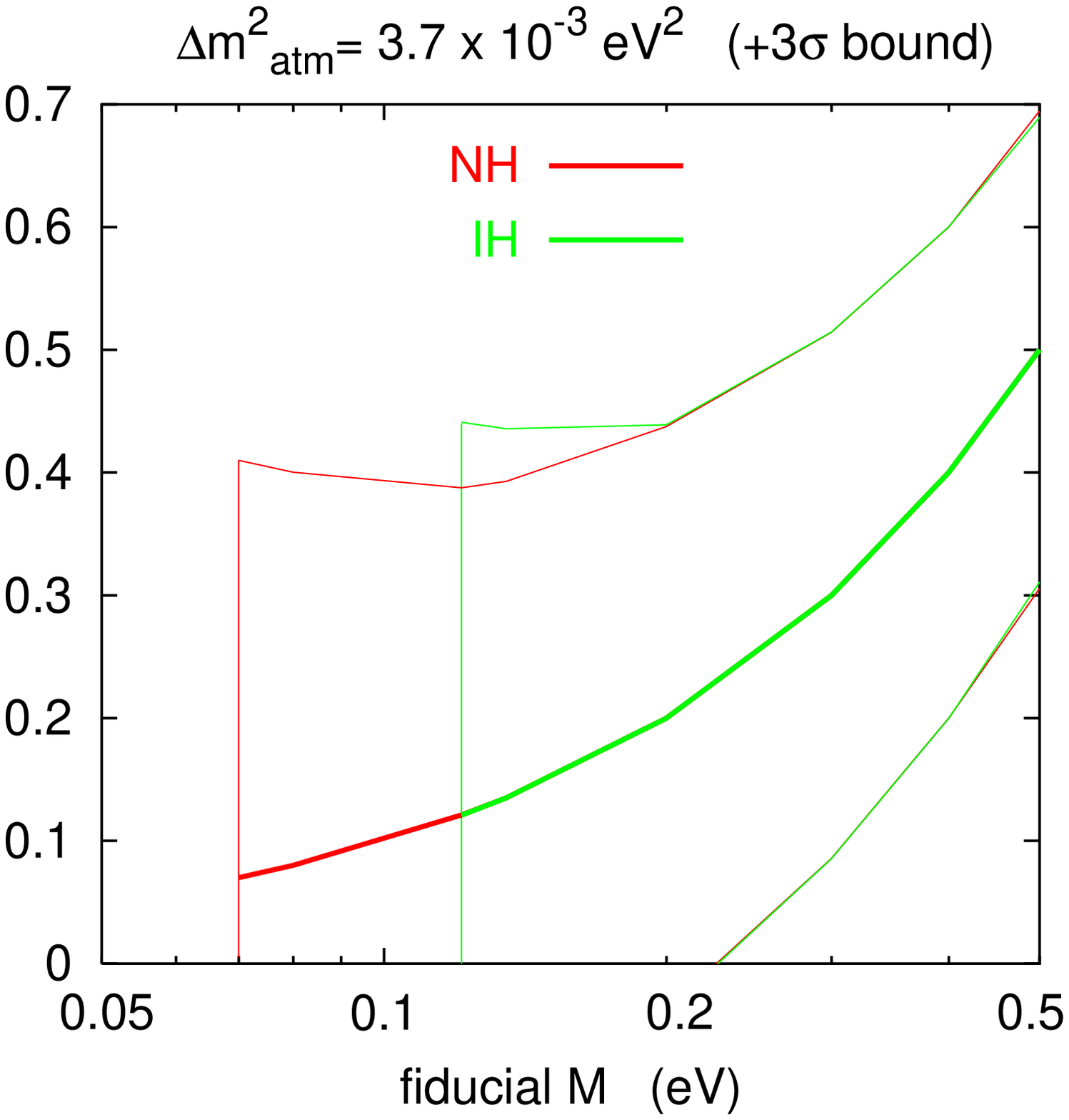}
\caption{\label{figPlanck} 
Predicted 2$\sigma$ error on the total neutrino mass $M \equiv \sum
m_\nu$ as a function of $M$ in the fiducial model, using PLANCK and
SDSS (limited to $k_{\rm max}=0.15~h$ Mpc$^{-1}$). The left plot was
obtained with the preferred experimental value of $\Delta m^2_{\rm
atm}$, and the right plot with the current 3$\sigma$ upper bound. In
each case, we show the results assuming either NH or IH.  
}
\end{figure}

We first derive the precision with which the combined PLANCK and SDSS
data will constrain the total neutrino mass in a near future.
Experimental specifications for these experiments are given in the
previous section, and we choose to limit SDSS data to the scale
$k_{\rm max}=0.15~h$ Mpc$^{-1}$ where non-linear effects are still
small.  Fig.\ \ref{figPlanck} shows the predicted 2$\sigma$ error on
$M$ for various fiducial models, assuming different values of $M$, the
two possible schemes for the mass splitting (either NH or IH), and two
different values of $\Delta m^2_{\rm atm}$. The solar mass scale
$\Delta m^2_{\rm sun}$ is essentially irrelevant in this analysis, and
is kept fixed to the current preferred value of $6.9 \times 10^{-5}$
eV$^2$. The possible values of $M$ are of course bounded from below:
the minimal value corresponds to the limit in which the lightest
neutrino mass goes to zero, in one of the two NH or IH schemes.

Let us first concentrate on the case in which $\Delta m^2_{\rm atm}$
has its current preferred value of $2.6 \times 10^{-3} {\rm eV}^2$
(left plot).  The minimal value of $M$ in the NH (resp.\ IH) case is
approximately 0.06 eV (resp.\ 0.10 eV). However, the 2$\sigma$
detection threshold, defined by $M= 2 \sigma(M)$, is around 0.21 eV.
We conclude that PLANCK+SDSS will probe mainly the region were the
three neutrinos are quasi-degenerate in mass, with no possibility to
distinguish between the two cases. In absence of clear detection, the
2$\sigma$ upper bound will be of order 0.2 eV, corresponding to
individual masses (0.08, 0.06, 0.06) eV assuming NH, or (0.073, 0.073,
0.053) eV assuming IH. As expected, we find that the
2$\sigma$ detection threshold is still 0.21 eV when the calculations
are performed with a larger value $\Delta m^2_{\rm atm}=3.7 \times
10^{-3}$ eV$^2$ (the 3$\sigma$ upper bound in Eq.\ \ref{dm2values}),
as shown in the right plot of Fig.\ \ref{figPlanck}.

It is interesting to study whether this precision is limited mainly by
a degeneracy between $M$ and some combination of other cosmological
parameters, or simply by the experimental sensitivity to the
individual effect of $M$. In the first case, the results could be
improved by including priors from other types of experiments on the
cosmological parameters; in the second case, one would have to wait
for a new generation of CMB and/or LSS experiments. In order to
address this point, we computed the eigenvectors and eigenvalues of
the reduced Fisher matrix. It turns out that for all our fiducial
models, one of the unit eigenvectors points precisely in the direction
of $M$, with coefficient very close to one in this direction (and, of
course, the corresponding eigenvalue matches the error previously
obtained for $M$). We conclude that $M$ is not affected by a parameter
degeneracy, and that independent measurements of other cosmological
parameters would not help very much in constraining neutrino masses.
Note that this is not yet the case for current
cosmological bounds on neutrino masses, where the addition of priors
on parameters such as the Hubble constant or $\Omega_{\Lambda}$
leads to more stringent bounds (see e.g.\ \cite{Crotty:2004gm}).

\begin{table}
\begin{ruledtabular}
\begin{tabular}{lccccccccccc}
& $\ln C^T_{200}$ & $n_s$ & $\tau$ & $\Omega_{\Lambda}$ & $\Omega_m h^2$ & 
$\Omega_b h^2$ & {\bf M (eV)} & $Y_{He}$ & $\ln[b^2 P(k_0)]$ 
& $X$ & \\
\\
\hline
9 parameters
& 0.005 & 0.007 & 0.005 & 0.01 & 0.001 & 0.0002 & {\bf 0.11} & 0.01 & 0.007 & --
\\   
~+ $X=N_{\nu}^{\rm r}$ & 0.005 & 0.008 & 0.005 & 0.01 & 0.003 & 0.0002 & {\bf 0.12} & 
0.01 & 0.007 & 0.14
\\
~+ $X=\Omega_{\rm k}$ & 0.005 & 0.008 & 0.005 & 0.01 & 0.002 & 0.0002 & {\bf 0.13} & 
0.01 & 0.007 & 0.003
\\
~+ $X=w$ & 0.005 & 0.008 & 0.005 & 0.01 & 0.002 & 0.0002 & {\bf 0.14} & 
0.01 & 0.007 & 0.05
\\
~+ $X=\alpha$ & 0.005 & 0.010 & 0.005 & 0.01 & 0.001 & 0.0002 & {\bf 0.11} & 
0.02 & 0.007 & 0.008
\end{tabular}
\end{ruledtabular}
\caption{\label{table_models}
Absolute errors at the 1-$\sigma$ level for various cosmological
models, using PLANCK+SDSS ($k_{\rm max}=0.15 \, h$ Mpc$^{-1}$).  The
first line shows our simplest flat $\Lambda$CDM model, described by 9
free parameters with fiducial values $C^T_{200}=0.85$, $n_s=0.96$,
$\tau=0.11$, $\Omega_{\Lambda}=0.70$, $\Omega_m h^2=0.143$, $\Omega_b
h^2=0.023$, $M=0.3~{\rm eV}$ (normal hierarchy), $Y_{He}=0.24$.
The value chosen for $b^2 P(k_0=0.1\, h$ Mpc$^{-1})$ is 
irrelevant.  The next lines have one additional parameter $X$: an
effective number of neutrinos $N_{\nu}^{\rm r}$ parametrizing the
abundance of extra relativistic relics, with fiducial value $0$; a free
spatial curvature parametrized by $\Omega_{\rm k}$ with fiducial value
$0$; a free time-independent equation of state for dark energy
parametrized by $w$ with fiducial value $-1$; a free scalar tilt
running parametrized by $\alpha=dn_s/d\ln k$ with fiducial value $0$.
}
\end{table}

\begin{table}
\begin{ruledtabular}
\begin{tabular}{lcccccccccccc}
&$k_{\rm max}~ (h/{\rm Mpc})$ & $\ln C^T_{200}$ & 
$n_s$ & $\tau$ & $\Omega_{\Lambda}$ & 
$\Omega_m h^2$ & $\Omega_b h^2$ & {\bf M (eV)} & 
$Y_{He}$ & $\ln[b^2 P(k_0)]$ 
& \\
\\
\hline
SDSS alone 
& 0.10 & -- & 0.6 & -- & 0.8 & 0.5 & 0.1 & {\bf 7.0} & -- & 0.3
\\
& 0.15 & -- & 0.5 & -- & 0.09 & 0.4 & 0.08 & {\bf 1.5} & -- & 0.06
\\
& 0.20 & -- & 0.1 & -- & 0.05 & 0.09 & 0.02 & {\bf 0.5} & -- & 0.01
\\
\hline
PLANCK (no pol.)& -- & 0.005 & 0.02 & 0.10 & 0.05 & 0.006 & 0.0006 & {\bf 0.42} & 
0.03 & --
\\
PLANCK (no pol.) + SDSS & 0.10 & 0.005 & 0.02 & 0.08 & 0.02 & 0.002 & 0.0004 & {\bf 0.24} & 
0.02 & 0.015
\\
& 0.15 & 0.005 & 0.02 & 0.08 & 0.01 & 0.001 & 0.0003 & {\bf 0.15} & 
0.02 & 0.008
\\
& 0.20 & 0.005 & 0.01 & 0.07 & 0.006 & 0.0009 & 0.0003 & {\bf 0.13} & 
0.02 & 0.005
\\   
\hline
PLANCK (all) & -- & 0.005 & 0.008 & 0.005 & 0.04 & 0.004 & 0.0003 & {\bf 0.30} & 
0.01 & --
\\
PLANCK (all) + SDSS & 0.10  & 0.005 & 0.007 & 0.005 & 0.02 & 0.002 & 0.0002 & {\bf 0.19} & 
0.01 & 0.012
\\
& 0.15 & 0.005 & 0.007 & 0.005 & 0.01 & 0.001 & 0.0002 & {\bf 0.11} & 
0.01 & 0.007
\\
& 0.20 & 0.004 & 0.007 & 0.005 & 0.006 & 0.0008 & 0.0002 & {\bf 0.08} & 
0.01 & 0.005
\\   
\hline
CMBpol & -- & 0.003 & 0.003 & 0.003 & 0.006 & 0.0006 & 0.00008 & {\bf 0.07} & 
0.004 & --
\\
CMBpol + SDSS & 0.10 & 0.003 & 0.003 & 0.003 & 0.006 & 0.0006 & 0.00008 & {\bf 0.07} & 
0.004 & 0.011
\\
& 0.15 & 0.003 & 0.003 & 0.003 & 0.005 & 0.0006 & 0.00007 & {\bf 0.06} & 
0.004 & 0.006
\\
& 0.20 & 0.003 & 0.003 & 0.003 & 0.004 & 0.0005 & 0.00007 & {\bf 0.05} & 
0.004 & 0.004
\end{tabular}
\end{ruledtabular}
\caption{\label{table_experiments}
Absolute errors at the 1-$\sigma$ level, for various experiments and
the same $\Lambda$CDM model as in table \ref{table_models} (with 9
free parameters). In particular, the fiducial value of the total
neutrino mass is still $M=0.3$ eV. When using SDSS, we show the
results for three choices of $k_{\rm max}$, the maximal wavenumber on
which the data are compared with linear theory predictions: $k_{\rm
  max}=0.10\, h$ Mpc$^{-1}$ (conservative), $0.15\, h$ Mpc$^{-1}$
(reasonable), or $0.20\, h$ Mpc$^{-1}$ (optimistic).}
\end{table}

\begin{table}
\begin{ruledtabular}
\begin{tabular}{lccc}
& SDSS & SDSS+KAOS & ``hypothetical LSS''
\\
\\
\hline
PLANCK & 0.21 & 0.16 & 0.11
\\   
CMBpol & 0.13 & 0.10 & 0.09
\\
``ideal CMB'' & 0.10 & 0.09 & 0.08
\end{tabular}
\end{ruledtabular}
\caption{\label{table_threshold}
2-$\sigma$ detection threshold (in eV) for various combinations of CMB
and LSS experiments (assuming the normal hierarchy scenario).  The
``ideal CMB'' experiment is limited only by cosmic variance up to
multipole $l=2500$ and covers 100\% of the sky. The ``hypothetical
LSS'' survey has a volume $V_{\rm eff} \simeq 40\,({\rm Gpc}/h)^3$ and
probes the linear spectrum up to $k_{\rm max} = 0.15\, h$ Mpc$^{-1}$
(that would be the case of a large galaxy survey covering 75\% of the
sky up to $z=0.8$).
}
\end{table}

The absence of large parameter degeneracies applies to our reference
model with nine free parameters. It may not
necessarily be true in the presence of extra parameters
describing deviations from the concordance $\Lambda$CDM model.  In
order to illustrate this point and to test the robustness of our
conclusions, we have calculated the error on each parameter for
several extended cosmological scenarios, with extra relativistic
degrees of freedom, spatial curvature, dark energy with varying
density but constant equation of state, or a primordial spectrum with
running tilt (see Table \ref{table_models}).  The neutrino mass bound
is found to be quite robust in all these cases, which proves that in
none of these models the effect of $M$ can be mimicked by some other
parameter combination.

It is also interesting to study the relative impact of CMB
temperature, CMB polarization and LSS data on the measurement of
$M$.  We show in table \ref{table_experiments} the error on each
parameter for SDSS alone, PLANCK alone (with or without polarization),
and various combinations of CMB and LSS data, with an explicit
dependence on the value of $k_{\rm max}$. The complementarity of
PLANCK and SDSS clearly appears. While PLANCK alone would achieve only
a 1$\sigma$ detection of $M=0.3$ eV and SDSS alone would not detect it
at all, the combined data would probe this value at the 3$\sigma$
level. One can check from Table \ref{table_experiments} that PLANCK
data on polarization lowers the error on $M$ by approximately $30\%$.
By diagonalizing the ``PLANCK (no pol.)+SDSS'' Fisher matrix, we
checked that without polarization there would be a significant
degeneracy between neutrino mass and optical depth to reionization.
Indeed, while reionization lowers the CMB temperature spectrum keeping
the matter power spectrum unchanged, the effect of neutrino
free-streaming is opposite in first approximation 
(at least on small scales). So,
polarization measurements are
indirectly a key ingredient for neutrino mass determination.
 
\subsection{Post-PLANCK experiments}

Here we consider whether future CMB and LSS experiments will reach a
better sensitivity on the neutrino mass, in particular at the level
of the small values of $M$ expected for the hierarchical normal and 
inverted schemes. Sensitivities significantly better than 0.1 eV would
mean approaching the absolute minimum of $M$ in the NH case or even
ruling out the IH scenario.

In the previous section, we mentioned a few CMB missions that have
been proposed so far in complement to PLANCK. We will study the impact
of a few of them, and of an ``ideal CMB experiment'' that would be
limited only by cosmic variance up to $l=2500$ (both for temperature
and polarization). The main difficulty for reaching this goal would be
to subtract accurately small-scale foregrounds, and in particular
point-like sources, but even with current technology such an ideal
experiment is conceivable. On the other hand, it is difficult to
specify the characteristics of an ideal LSS experiment, since it will
be limited by technological improvements in instrumentation and data
processing. Therefore, we will keep in the analysis a free parameter
$V_{\rm eff}$ describing the effective volume of an ideal
volume-limited survey.

\begin{figure}
\includegraphics[width=.85\textwidth]{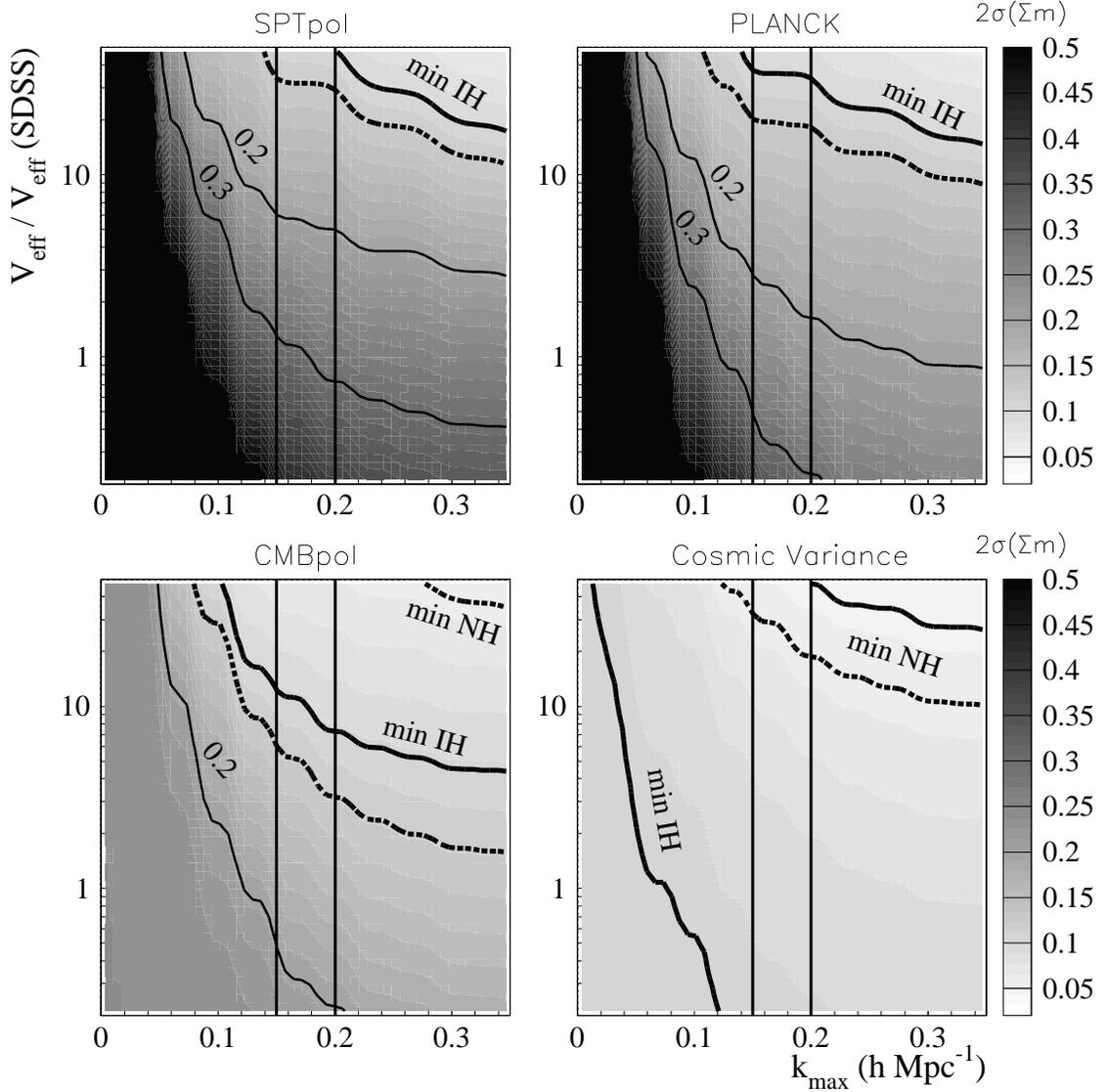}
\caption{\label{figall}
  The grey regions are the 2$\sigma$ expected errors on $\sum m_\nu$
  (eV) for a fiducial value of 0.11 eV, as a function of the
  parameters of the galaxy survey, where each panel corresponds to a
  specific CMB experiment.  The vertical lines indicate the cut-off
  wavenumber $k_{\rm max}$ for the linear matter power spectrum at the
  conservative (optimistic) value $0.15(0.2)\, h$ Mpc$^{-1}$. The thin
  contours shown are (from bottom to top) for 0.3 and 0.2 eV, while
  the thick contours correspond to the minimum values of $\sum m_\nu$
  in the IH (lower lines) and NH (upper lines) schemes, assuming the
  best-fit (solid) or the 3$\sigma$ upper bound (dashed) value of
  $\Delta m^2_{\rm atm}$.  }
\end{figure}

We show in Fig.\ \ref{figall} the predicted 2$\sigma$ error in four
cases corresponding to SPTpol (upper left), PLANCK (upper right),
CMBpol (lower left), and our ideal CMB experiment (lower right).  The
value of 2$\sigma$ (in eV) is shown with grey levels, as a function of
$k_{\rm max}$ (horizontal axis) and $V_{\rm eff}$ (vertical axis) in
units of $V_{\rm eff}({\rm SDSS})=1~({\rm Gpc}/h)^3$.
The total mass has been fixed to $M=0.11$ eV, distributed according
to the NH scheme. We learned from the previous subsection that for
higher values of $M$, the error could be smaller (at most by a factor
2). However, we are now interested in the range $0.05~{\rm eV} < M <
0.2~{\rm eV}$, since larger values should be detected by PLANCK+SDSS,
and smaller values are excluded by oscillation experiments. In this
range, on can safely interpolate the results obtained at $M=0.11$ eV.
In particular, our results for a cosmic-variance limited
CMB experiment are in reasonable agreement with those of
\cite{Hannestad:2002cn}.

For SDSS (or for any survey with $z<1$) we expect the relevant value
of $k_{\rm max}$ to be around $0.15\, h$ Mpc$^{-1}$. However,
depending on the overall amplitude of the matter power spectrum (often
parametrized by $\sigma_8$, and still poorly constrained) and on
future improvements in our understanding of non-linear corrections,
this value might appear to be either too optimistic or too
pessimistic: this is the reason why it is interesting to leave it as a
free parameter.

One can see that replacing PLANCK by CMBpol would lead to a better
sensitivity to the neutrino mass, with a 2$\sigma$ detection threshold
at 0.13 eV instead of 0.21 eV.  The expected errors for CMBpol, with
and without SDSS data, can be found in Table \ref{table_experiments}.
Adding to SDSS the two KAOS surveys (centered around $z=1$ and $z=3$)
would also lead to some improvement. For Planck+SDSS+KAOS we get a
2$\sigma$ detection threshold of $M \sim 0.16$ eV, while for
CMBpol+SDSS+KAOS one could reach $M \sim 0.10$ eV.  These results are
summarized in Table \ref{table_threshold}.

\begin{figure}
\includegraphics[width=.48\textwidth]{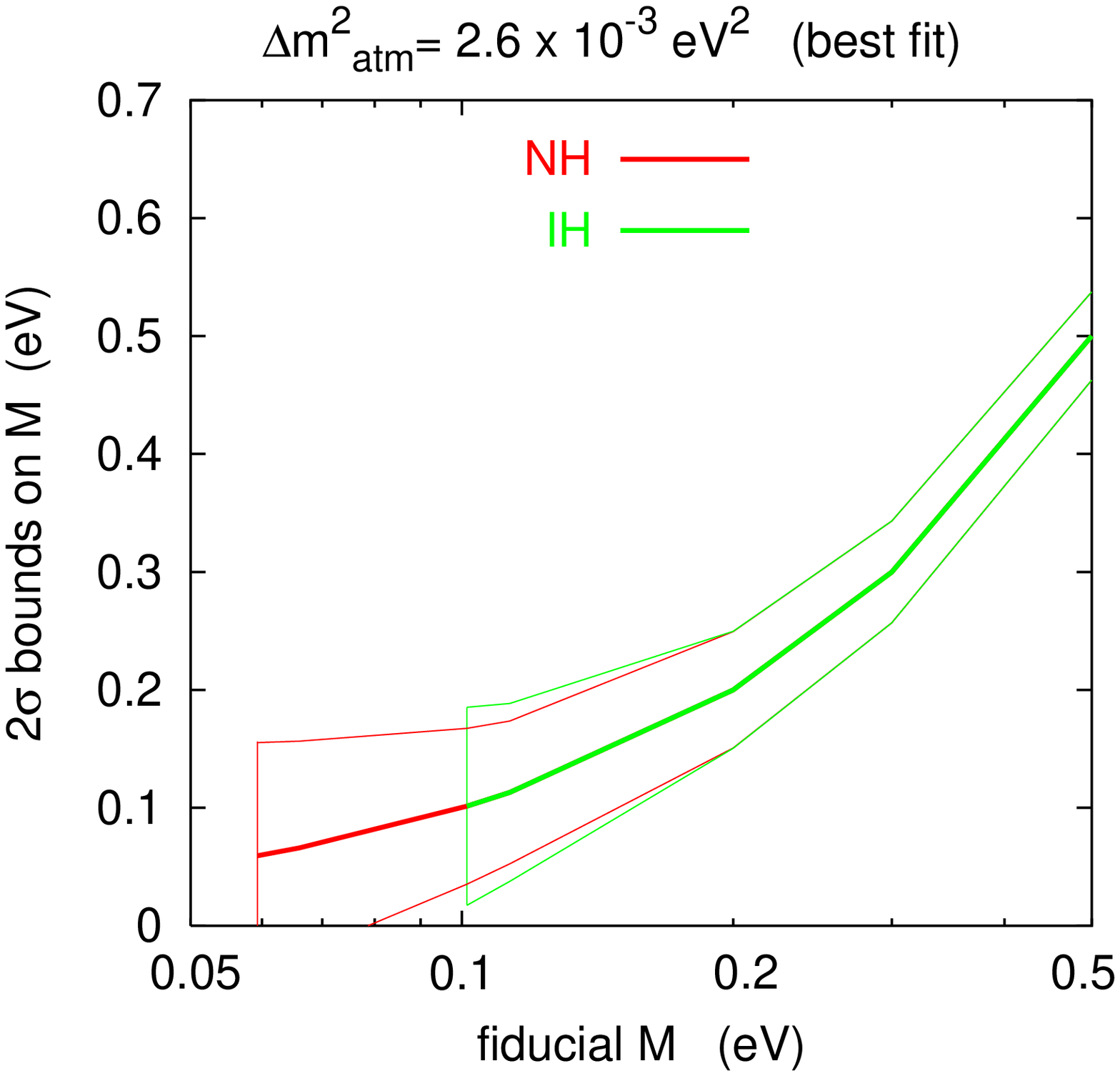}
\includegraphics[width=.48\textwidth]{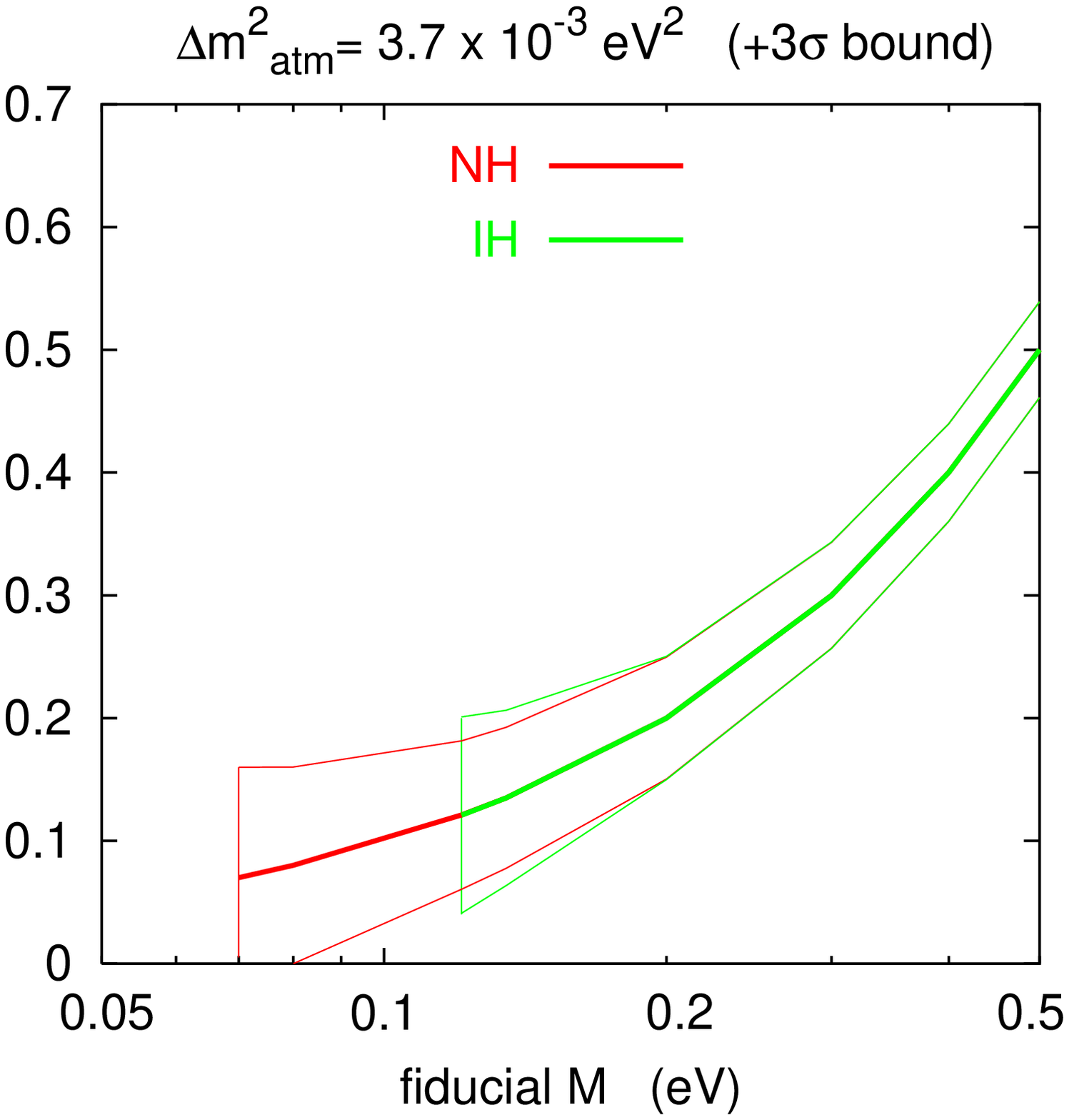}
\caption{\label{figIdeal}
  Predicted 2$\sigma$ error on $\sum m_\nu$ as a function of $\sum m_\nu$ 
  in the
  fiducial model, using an ideal CMB experiment (limited only by cosmic 
  variance up to $l=2500$, both for temperature and polarization) and 
  a redshift survey covering 75\% of the sky up to $z \simeq 0.8$
  ($V_{\rm eff} = 40~({\rm Gpc}/h)^3$), still limited to 
  $k_{\rm max}=0.15\, h$ Mpc$^{-1}$. The left plot was obtained with the
  preferred experimental value of $\Delta m^2_{\rm atm}$, and the right 
  plot with the current 3$\sigma$ upper bound. In each case, we show the
  results assuming either NH or IH.
}
\end{figure}

There is still room for improvement beyond this set of experiments. In
order to make a precise statement on the conclusions that could be
drawn on the long term, we keep the ``ideal CMB experiment''
characteristics and fix $V_{\rm eff}$ to $40~({\rm Gpc}/h)^3$ (in
section \ref{sec:futureCMBandLSS}, we argued that this could hopefully
represent the volume of a survey comparable to the LSST project),
while keeping $k_{\rm max} = 0.15\, h$ Mpc$^{-1}$.  In Fig.\ 
\ref{figIdeal}, we plot the corresponding results in the same way as
we did for PLANCK+SDSS.  Assuming the IH scenario, we see that any
value of the mass could be detected at the 2$\sigma$ level.  Assuming
NH, this is only true at the 1 or 1.5$\sigma$ level, depending on the
value of $\Delta m^2_{\rm atm}$. The 2$\sigma$ detection threshold is
at 0.08 eV.

Our results show, for the first time, that if the available
cosmological data are precise enough, the expected errors on the
neutrino masses depend not only on the sum of neutrino masses, but
also on what is assumed for the mass splitting between the neutrino
states. As can be seen from Figs.\ \ref{figPlanck} and \ref{figIdeal},
the sensitivity on $M$ will be slightly better in the NH case in the
mass region close to the minimum value of the IH scheme. These small
differences arise from the changes in the free-streaming effect that
we have described in section \ref{sec:numasses}, and obviously
disappear for a total mass in the quasi-degenerate region (above 0.2
eV or so).

In any case, the main contribution of cosmology to the possible
discrimination between the neutrino mass schemes will still be the
possibility of ruling out the case in which the masses are
quasi-degenerate. Even in our most optimistic forecast (Fig.\ 
\ref{figIdeal}), if the preferred value of $M$ turns out to be smaller
than 0.1 eV, the error bar will still be too large in order to safely
rule out the IH case. We also performed an extended analysis in which,
instead of assuming either normal or inverted hierarchy, we introduced
a tenth free parameter accounting for a continuous interpolation of
the mass spectrum between the two scenarios, for fixed $M$.  By
computing the error on this parameter, we obtained a confirmation that
the NH and IH scenarios cannot be discriminated directly from the
data.  However, any analysis of future, very precise cosmological data
must take into account the texture of neutrino masses in order to
translate the corresponding positive signal (or bound) into $M$.

\section{Conclusions}

In this paper we have analyzed the sensitivities of future CMB and LSS
data to the absolute scale of neutrino masses, taking into account
realistic experimental sensitivities and extending the results of
previous works
\cite{Hu:1997mj,Eisenstein:1998hr,Lesgourgues:1999ej,Hannestad:2002cn}.

We have considered the values of neutrino masses distributed according
to the presently favored three neutrino mass schemes, that follow
either a normal or an inverted hierarchy. As discussed in section
\ref{sec:numasses}, a different distribution of the same total
neutrino mass leads to small changes in the cosmological evolution of
neutrinos, and in particular in the free-streaming scales
(qualitatively discussed, for instance, in
\cite{Hu:1997mj,Hannestad:2002cn}). These changes disappear when the
total neutrino mass enters the quasi-degenerate region.

We used the Fisher matrix method to forecast the errors on
cosmological parameters that can be extracted from future
CMB experiment and redshift survey data, assuming a fiducial
9-dimensional cosmological model close to the currently favored
$\Lambda$CDM model. Our theoretical CMB and matter power spectra were
generated with the standard Boltzmann code CMBFAST, modified in order
to include three neutrino states with different masses.

In particular, for the case of PLANCK and SDSS we found good agreement
with the results of \cite{Eisenstein:1998hr}, with a 2$\sigma$-error
on the total neutrino mass of $0.2$ eV that will allow us to probe
only the quasi-degenerate neutrino mass region. Better sensitivity
will be achieved with the combination CMBpol and SDSS, for which we
found $0.12$ eV, close to the minimum value of the total neutrino mass
in the inverted hierarchy case. These results correspond to a
conservative value of $k_{\rm max} = 0.15\, h$ Mpc$^{-1}$, the maximal
wavenumber on which the LSS data are compared with the predictions of
linear theory. We also tested how the errors change when including
additional cosmological parameters to our fiducial model. In general,
we found that the errors on the neutrino masses are not modified in a
significant way.

Our results show that the approach where CMB experiments are only
limited by cosmic variance (as in \cite{Hannestad:2002cn}) is probably
too simplistic. However, if a future CMB experiment is capable of
getting close to such an ideal limit, then the combination with data
from galaxy redshift surveys larger than SDSS would lead to errors on
the total neutrino mass comparable to the minimum values of the
hierarchical scenarios. In such a case, we have shown that there exist
slight differences in the expected errors between the two hierarchical
neutrino schemes for the same total neutrino mass.

In conclusion, we consider that cosmological data can provide valuable
information on the absolute scale of neutrino masses, that nicely
complements the present and future projects of beta decay and
neutrinoless double beta decay experiments.  This conclusion is
reinforced when one takes into account other cosmological probes of
neutrino masses, complementary to the approach of the present paper.
We can cite, for instance, studies of the distribution of matter in the
Universe through the distortions of CMB maps caused by gravitational
lensing (measured from non-gaussianities in the CMB maps) 
\cite{Kaplinghat:2003bh}
and the weak gravitational lensing of background galaxies by
intervening matter \cite{Hu:1999ek,Hu:2002rm,Abazajian:2002ck,Song:2003gg}.

It is interesting to note that any information on the absolute
neutrino masses from cosmology will be interesting not only for
theoretical neutrino models, but also for connected baryogenesis
scenarios which occur through a leptogenesis process 
(see e.g.\ \cite{Giudice:2003jh,Hambye:2003rt,Buchmuller:2004nz}).

\section*{Acknowledgments}
We thank Ilenia Picardi for initial discussions concerning the present
work, as well as Martin Hirsch and Simon Prunet for various
suggestions, and Bruce Bassett for pointing us to the KAOS proposal.
This research was supported by a CICYT-IN2P3 agreement. SP was
supported by the Spanish grant BFM2002-00345, the ESF network Neutrino
Astrophysics and a Ram\'on y Cajal contract of MCyT.


\begin{thebibliography}{99}

\bibitem{Bonn:tw}
J.~Bonn et al.,
Nucl.\ Phys.\ Proc.\ Suppl.\  {\bf 91}, 273 (2001).

\bibitem{Osipowicz:2001sq}
A.~Osipowicz et al.\  [KATRIN Coll.],
hep-ex/0109033.

\bibitem{Elliott:2002xe}
S.R.~Elliott and P.~Vogel,
Ann.\ Rev.\ Nucl.\ Part.\ Sci.\  {\bf 52} 115, (2002)
[hep-ph/0202264].

\bibitem{Dolgov:2002wy}
A.D.~Dolgov,
Phys.\ Rept.\  {\bf 370}, 333 (2002)
[hep-ph/0202122].

\bibitem{Hu:1997mj}
W.~Hu, D.J.~Eisenstein and M.~Tegmark,
Phys.\ Rev.\ Lett.\  {\bf 80}, 5255 (1998) 
[astro-ph/9712057].

\bibitem{Spergel:2003cb}
D.N.~Spergel et al.,
Astrophys.\ J.\ Suppl.\  {\bf 148}, 175 (2003)
[astro-ph/0302209].

\bibitem{Hannestad:2003xv}
S.~Hannestad,
JCAP {\bf 0305}, 004 (2003)
[astro-ph/0303076].

\bibitem{Elgaroy:2003yh}
\O.~Elgar\o y and O.~Lahav,
JCAP {\bf 0304}, 004 (2003)
[astro-ph/0303089].

\bibitem{Tegmark:2003ud}
M.~Tegmark et al.\  [SDSS Coll.],
Phys.\ Rev.\ D, in press [astro-ph/0310723].

\bibitem{Barger:2003vs}
V.~Barger, D.~Marfatia and A.~Tregre,
hep-ph/0312065.

\bibitem{Hannestad:2003ye}
S.~Hannestad and G.~Raffelt,
hep-ph/0312154.

\bibitem{Crotty:2004gm}
P.~Crotty, J.~Lesgourgues and S.~Pastor,
hep-ph/0402049.

\bibitem{Eisenstein:1998hr}
D.J.~Eisenstein, W.~Hu and M.~Tegmark,
Astrophys.\ J.\  {\bf 518}, 2 (1998)
[astro-ph/9807130].

\bibitem{Lesgourgues:1999ej}
J.~Lesgourgues, S.~Pastor and S.~Prunet,
Phys.\ Rev.\ D {\bf 62}, 023001 (2000)
[hep-ph/9912363].

\bibitem{Hannestad:2002cn}
S.~Hannestad,
Phys.\ Rev.\ D {\bf 67}, 085017 (2003) 
[astro-ph/0211106].

\bibitem{Abazajian:2002ck}
K.N.~Abazajian and S.~Dodelson,
Phys.\ Rev.\ Lett.\  {\bf 91}, 041301 (2003) 
[astro-ph/0212216].

\bibitem{Kaplinghat:2003bh}
M.~Kaplinghat, L.~Knox and Y.S.~Song,
Phys.\ Rev.\ Lett.\  {\bf 91}, 241301 (2003) 
[astro-ph/0303344].

\bibitem{Maltoni:2003da}
M.~Maltoni, T.~Schwetz, M.A.~T\'ortola and J.W.F.~Valle,
Phys.\ Rev.\ D {\bf 68}, 113010 (2003)
[hep-ph/0309130].

\bibitem{Fogli:2003th}
G.L.~Fogli, E.~Lisi, A.~Marrone and D.~Montanino,
Phys.\ Rev.\ D {\bf 67}, 093006 (2003)
[hep-ph/0303064].

\bibitem{Gonzalez-Garcia:2003qf}
M.C.~Gonz\'alez-Garc\'{\i}a and C.~Pe\~na-Garay,
Phys.\ Rev.\ D {\bf 68} (2003) 093003
[hep-ph/0306001].

\bibitem{Huber:2004ug}
P.~Huber, M.~Lindner, M.~Rolinec, T.~Schwetz and W.~Winter,
hep-ph/0403068.

\bibitem{Aguilar:2001ty}
A.~Aguilar et al.\  [LSND Coll.],
Phys.\ Rev.\ D {\bf 64} (2001) 112007
[hep-ex/0104049].

\bibitem{Gonzalez-Garcia:2002dz}
M.C.~Gonz\'alez-Garc\'{\i}a and Y.~Nir,
Rev.\ Mod.\ Phys.\  {\bf 75}, 345 (2003)
[hep-ph/0202058].
 
\bibitem{Barger:2003qi}
V.~Barger, D.~Marfatia and K.~Whisnant,
Int.\ J.\ Mod.\ Phys.\ E {\bf 12}, 569 (2003)
[hep-ph/0308123].

\bibitem{Dolgov:1997mb}
A.D.~Dolgov, S.H.~Hansen and D.V.~Semikoz,
Nucl.\ Phys.\ B {\bf 503}, 426 (1997)
[hep-ph/9703315].

\bibitem{Mangano:2001iu}
G.~Mangano, G.~Miele, S.~Pastor and M.~Peloso,
Phys.\ Lett.\ B {\bf 534}, 8 (2002)
[astro-ph/0111408].

\bibitem{Dolgov:2002ab}
A.D.~Dolgov et al.,
Nucl.\ Phys.\ B {\bf 632}, 363 (2002)
[hep-ph/0201287].

\bibitem{Lewis:2002nc}
A.~Lewis and A.~Challinor,
Phys.\ Rev.\ D {\bf 66}, 023531 (2002)
[astro-ph/0203507].

\bibitem{Primack:1994pe}
J.R.~Primack, J.~Holtzman, A.~Klypin and D.O.~Caldwell,
Phys.\ Rev.\ Lett.\  {\bf 74}, 2160 (1995)
[astro-ph/9411020].

\bibitem{Seljak:1996is}
U.~Seljak and M.~Zaldarriaga,
Astrophys.\ J.\  {\bf 469}, 437 (1996)
[astro-ph/9603033].

\bibitem{Goldstein:2002gf}
J.H.~Goldstein et al.,
Astrophys.\ J.\  {\bf 599}, 773 (2003)
[astro-ph/0212517].

\bibitem{Readhead:2004gy}
A.C.S.~Readhead et al.,
astro-ph/0402359.

\bibitem{Rebolo:2004vp}
R.~Rebolo et al.,
astro-ph/0402466.

\bibitem{Bowden:2003ub}
M.~Bowden et al.,
astro-ph/0309610.

\bibitem{Tegmark:1999ke}
M.~Tegmark, D.J.~Eisenstein, W.~Hu and A.~de Oliveira-Costa,
Astrophys.\ J.\  {\bf 530}, 133 (2000)
[astro-ph/9905257].

\bibitem{Prunet:1999fj}
S.~Prunet, S.K.~Sethi and F.R.~Bouchet,
Mon.\ Not.\ Roy.\ Astron.\ Soc.\  {\bf 314}, 358 (2000)
[astro-ph/9911243].

\bibitem{Patanchon:2003dg}
G.~Patanchon, H.~Snoussi, J.F.~Cardoso and J.~Delabrouille,
astro-ph/0302078.

\bibitem{Hu:2001fb}
W.~Hu,
Phys.\ Rev.\ D {\bf 65}, 023003 (2002)
[astro-ph/0108090].

\bibitem{Hu:2001kj}
W.~Hu and T.~Okamoto,
Astrophys.\ J.\  {\bf 574}, 566 (2002)
[astro-ph/0111606].

\bibitem{Okamoto:2003zw}
T.~Okamoto and W.~Hu,
Phys.\ Rev.\ D {\bf 67}, 083002 (2003)
[astro-ph/0301031].

\bibitem{Tegmark:1997rp}
M.~Tegmark,
Phys.\ Rev.\ Lett.\  {\bf 79}, 3806 (1997) 
[astro-ph/9706198].

\bibitem{Tyson:2003kb}
J.A.~Tyson  [the LSST Coll.],
Proc.\ SPIE Int.\ Soc.\ Opt.\ Eng.\  {\bf 4836}, 10 (2002)
[astro-ph/0302102].

\bibitem{Lopez:1998aq}
R.E.~Lopez, S.~Dodelson, A.~Heckler and M.S.~Turner,
Phys.\ Rev.\ Lett.\  {\bf 82}, 3952 (1999)
[astro-ph/9803095].

\bibitem{Bowen:2001in}
R.~Bowen et al.,
Mon.\ Not.\ Roy.\ Astron.\ Soc.\  {\bf 334}, 760 (2002)
[astro-ph/0110636].

\bibitem{Bashinsky:2003tk}
S.~Bashinsky and U.~Seljak,
Phys.\ Rev.\ D, in press [astro-ph/0310198].

\bibitem{Cuoco:2003cu}
A.~Cuoco, F.~Iocco, G.~Mangano, G.~Miele, O.~Pisanti and P.D.~Serpico,
astro-ph/0307213.

\bibitem{Cyburt:2003fe}
R.H.~Cyburt, B.D.~Fields and K.A.~Olive,
Phys.\ Lett.\ B {\bf 567}, 227 (2003)
[astro-ph/0302431].


\bibitem{Jungman:1995av}
G.~Jungman, M.~Kamionkowski, A.~Kosowsky and D.N.~Spergel,
Phys.\ Rev.\ Lett.\  {\bf 76}, 1007 (1996)
[astro-ph/9507080].


\bibitem{Song:2003gg}
Y.S.~Song and L.~Knox,
astro-ph/0312175.

\bibitem{Hu:1999ek}
W.~Hu,
Astrophys.\ J.\  {\bf 522}, L21 (1999)
[astro-ph/9904153].

\bibitem{Hu:2002rm}
W.~Hu,
Phys.\ Rev.\ D {\bf 66}, 083515 (2002)
[astro-ph/0208093].

\bibitem{Giudice:2003jh}
G.F.~Giudice, A.~Notari, M.~Raidal, A.~Riotto and A.~Strumia,
hep-ph/0310123.

\bibitem{Hambye:2003rt}
T.~Hambye, Y.~Lin, A.~Notari, M.~Papucci and A.~Strumia,
hep-ph/0312203.

\bibitem{Buchmuller:2004nz}
W.~Buchm\"uller, P.~Di Bari and M.~Pl\"umacher,
hep-ph/0401240.


\end{thebibliography}
\end{document}